\documentstyle[aps,prl,epsf]{revtex}  

\begin{document}

\twocolumn[\hsize\textwidth\columnwidth\hsize\csname
@twocolumnfalse\endcsname

\title{Boundary lubrication properties of materials with expansive freezing}
\author{E. A. Jagla}
\address{Centro At\'omico Bariloche and Instituto Balseiro,
Comisi\'on Nacional de Energ\'{\i}a At\'omica\\
(8400) S. C. de Bariloche, Argentina}
\maketitle
\begin{abstract}
 
We have performed molecular dynamics simulations of solid-solid
contacts lubricated by a model fluid displaying many of the properties
of water, particularly its expansive freezing.
Near the region where expansive freezing occurs, 
the lubricating film remains fluid, and the friction force 
decreases linearly as the shear velocity is reduced. No sign of stick-slip 
motion is observed even at the lowest velocities.
We give a simple interpretation of these results,
and suggest that in general good boundary 
lubrication properties will be
found in the family of materials with expansive freezing.

\end{abstract}

\vskip2pc] \narrowtext

The understanding from a microscopic point of view of the tribological properties
of materials in the boundary lubrication regime 
is of great importance for the
technological development of new lubricants for new applications\cite{1}.
Micro-mechanical devices and high density magnetic storage media are just
two examples where the detailed properties of the lubricants may affect in 
an important way the performance of the system. 
To attain low friction coefficients with lubricating films 
a few molecular layers thick is just the most obvious goal of lubricant 
design.

Fluid lubrication is a widespread strategy to reduce friction.
However it suffers of some limitations in the boundary regime \cite{4}, 
where the fluid is
only a few molecular layers thick.
Fluids have a general tendency to  attain a solid-like structure when 
squeezed between two solid surfaces\cite{4,2}. This confinement-induced
solidification is responsible for microscopic stick-slip motion (SSM)\cite{4} when
the confining surfaces are sheared past each other.
SSM in turn generates a finite friction 
force in the limit of
low velocities, and then a non-zero friction coefficient.
In fact, mechanisms that prevent SSM have been seen to reduce 
dramatically the friction force between the lubricated surfaces\cite{21,21b}.

In the boundary regime a lubricant composed of complex, branch chain
molecules usually performs better 
than one formed by spherical, or single chain molecules, 
as the former does not readily arrange in a solid-like structure\cite{5},
and allows for a smooth 
shearing of the surfaces, in a state of very low friction.
In addition, fluids with branch chain molecules have typically 
higher bulk viscosity than those with spherical, or single chain molecules.
This justifies the experimentally 
observed (see Table 1 in Ref. \cite{5})
negative correlation existing between bulk viscosity
and friction coefficient under boundary lubrication conditions.

Water is a noteworthy exception to this rule \cite{5}.
The water molecule is a small, almost spherical particle, and the viscosity 
of water is very low. 
However, carefully controlled experiments using the surface force apparatus
show that water is an excellent boundary
lubricant at a microscopic scale\cite{6,7,8}. In fact, 
the friction coefficient between
two water lubricated, atomically flat mica surfaces  was found\cite{5,6}
to be around 0.01, in the range
of the lowest values obtained with other fluids 
in the boundary regime.
The reason for the anomalously good lubrication properties of water 
has not been addressed in detail. 
In this paper we show by studying a model system, 
that the good lubrication properties of
water can be associated to its expansive freezing, and that good boundary lubricants
can be fetched in general within the class of materials having this property.

Many of the anomalous properties of water (and tetrahedrally coordinated 
liquids in general\cite{9}) have been recently systematized by the use of a model 
system of spherical particles interacting through a soft-core potential
\cite{10,11,12}. Its main characteristic is the existence of two stable relative 
distances between particles, as a function of the applied pressure. 
It has been suggested that these two distances are realized in water 
owing to two distinct arrangements of hydrogen bonds\cite{13}. We use one such 
model\cite{10} to simulate a lubricant film. The interaction potential as a 
function of the interparticle distance $r$ has a strict hard core at a 
distance $r=\sigma$ and a repulsive ramp extending from $\sigma$  to $\alpha\sigma$ 
($\alpha>1$, $\alpha=1.5$ hereafter), 
in which the potential energy changes linearly from $\varepsilon$ to zero. 
Natural units for this problem 
are the length $\sigma$, the energy $\varepsilon$, and the mass of the particles 
$m$, from 
which a time scale $\tau\equiv\sqrt{m\sigma^2/\varepsilon}$ is defined.

The bulk system freezes into a closed packed lattice with lattice 
parameter $\sim\alpha\sigma$  
at low temperatures and low pressures. The melting curve of 
this crystalline structure was determined previously by Monte Carlo 
simulations\cite{10}, and is reproduced in Fig. 1(a) for completeness. The melting 
temperature increases as a function of pressure at the lowest values of 
$P$ and $T$, consistent with the fact that in this region particles behave 
as hard spheres of radius $\alpha\sigma/2$. At higher pressures, the melting line 
reenters toward lower temperatures as neighbor particles can get closer 
than  $\alpha\sigma$\cite{14}. Other crystalline structures exist at 
higher pressures\cite{14,15}.

The geometrical arrangement we use for the simulations is similar to that used
in Refs. [17] and [18], and is sketched in Fig. \ref{f0}(b). It qualitatively 
models the configuration used in the surface force apparatus \cite{1}.
Two opposing pieces of crystalline fcc solids exposing (111) faces
are lubricated by our model system.
Particles within the solids are assumed to be fixed to their lattice 
positions, and the interactions between particles of the fluid and those 
of the solids are taken equal to the intra-fluid interactions. 
The lattice parameter of the solid blocks 
$a_0$ is chosen for each pressure as that of the bulk solid at the melting 
temperature. This is not a crucial choice, and is made only to avoid
introducing a bias between the different simulated points.
The lubricant particles within the gap between the solids are at dynamic
equilibrium with lubricant particles outside the gap.
For the static (no shear) simulations, periodic boundary conditions
are applied in the three directions. The solid blocks are of finite
size $l_x$ in the $x$-direction, and extend over the whole length of the 
simulation cell $l_y$ in the $y$-direction. The size of the simulation box
in the $x$-direction $\tilde {l_x}$ is substantially larger than $l_x$, 
allowing to reach a dynamical equilibrium between the lubricant inside the
gap, and the bulk fluid outside it.
$\tilde {l_x}$ is allowed to change according to the value of a hydrostatic 
external pressure 
$P$. Real fluids always possess some attraction between particles,
that we have not included. It is known however that the
attraction between particles (both fluid-fluid and fluid-solid)
in the van der Waals limit can be readily included in the results obtained
without attraction by the renormalization of the external pressure $P$
\cite{10}. Our hydrostatic pressure $P$ accounts for the attraction
between particles in the van der Waals limit, this attraction
being of the same strength
between fluid-fluid particles, and fluid-solid ones. Other cases remain
to be studied.

We have chosen $l_x\times l_y$  equal to $15\times 15 \sigma^2$. 
Each lubricant monolayer within the gap has about 110 particles,
and the total number of fluid particles 
simulated is 800. We have performed checks with systems of up to 
3500 fluid particles and $l_x\times l_y$  equal to $30\times 30 \sigma^2$, 
to assure that size effects do not spoil the
results.
We simulated systems at points {\bf A} and {\bf B}, 
indicated in Fig. 
1 ($P_A=0.1~\varepsilon/\sigma^3$, $T_A=0.035~\varepsilon /k_B$, 
$P_B=0.9~\varepsilon/\sigma^3$, $T_B=0.13~\varepsilon /k_B$). 
We calculated the normal force exerted by the 
lubricant --the solvation force-- and studied the structure of the film as a 
function of the distance $d$ between the two solid surfaces. In Fig. 2
we show results at point {\bf A}, where
the system behaves basically as a hard sphere system. 
When $d$ is only a few molecular 
radii, the solvation 
force per unit area $F_s$  shows a well known\cite{1,3,16,17,a1,a2,a3} oscillatory behavior 
(centered at the value of the hydrostatic pressure $P$)
associated to the layering of the fluid. The number of particles 
inside the gap $n_p$ (which is normalized to one
for a full layer of particles) has a staircase-like 
structure, remaining approximately constant as $F_s$ increases (we describe 
the process as $d$ is reduced), and jumping down coincidentally with the 
reduction of $F_s$, when a layer of lubricant  is squeezed out of the gap. 
Gray regions in Fig. 2 indicate precisely these squeezing-out regions,
where the number of layers is changing.
We measure the horizontal order within the lubricant through the structure 
factor $f$ of the layers adjacent to the solids (calculated at $|k|=4\pi/a_0$), 
and the two dimensional (horizontal) diffusivity of 
particles in the gap $D$. Within a plateau 
of $n_p$, $f$ increases and $D$ decreases as the film is compressed, indicating 
that the film becomes more structured as $F_s$ increases. In the squeezing-out
regions (the gray regions) $f$  
drops and $D$ maximizes, indicating that the film becomes less structured 
and more fluid. This whole picture is consistent with previous 
studies\cite{3,16}, and it can also be given a straightforward interpretation: 
if we assimilate $F_s$ to a sort of local pressure in the film, the 
increasing of order as $F_s$ increases is consistent with the fact that for 
the bulk material positional order increases and diffusivity decreases 
with the applied pressure. When a layer is being squeezed out, more 
space is available to the remaining particles that then attain a more 
fluid configuration.

Results are shown in Fig. 3 near the region where expansive freezing
occurs (point {\bf B} in Fig. 1). $F_s$ shows the same oscillating behavior 
as before. However, the structure of the film is now qualitatively 
different. The film remains in a liquid-like state as $F_s$ increases 
(as seen from the values of $f$ and $D$). This occurs because particles, 
being at distance $\sim~ 1.5 \sigma$  from their neighbors, can go closer 
(up to distance $\sim \sigma$) 
upon a reduction of $d$, without generating a higher degree of order in 
the system. Only after $F_s$ decreases the film becomes more solid-like 
(with higher $f$ and lower $D$). This is also consistent with the fact 
that this kind of model (as well as real water\cite{18}) is known to have 
a range of pressure in which diffusivity increases as pressure 
increases\cite{19}. Close to this region (which occurs near the maximum of 
the $P(T)$ melting line), a positive correlation between the values of  
$F_s$ and $D$ can be expected.

The inverted alternate at point {\bf B} of liquid-like and solid-like 
structure of the film upon compression represents a crucial difference 
from an experimental point of view, since typically a condition of 
constant load is applied to the contact. 
For stability reasons, this external load 
will be sustained by a lubricating film at a point with negative 
derivative of $F_s$  with respect to $d$, namely, it will be a solid-like 
film at point {\bf A}, and a liquid-like film at point {\bf B}. 
As we will see, this difference has profound consequences in the lubricating 
properties of the system.

Shearing simulations were performed in the same geometry than that 
used for the previous calculations, but now the upper solid block 
(which is now allowed to move in the $y$ direction) is given a mass $M$ 
($M=450~m$), and is attached to a horizontal spring (spring constant 
$k=0.45~\varepsilon/\sigma^2$) whose free 
end is moved in the $y$ direction at a constant velocity $v$ (see also [4]).
An external 
load  per unit area $L$ of $0.05~\varepsilon/\sigma^3$ 
in excess of the hydrostatic pressure $P$ within 
the lubricant is applied to the block along the $z$ direction, and the 
distance $d$ is allowed to change dynamically during the simulation.

Both at points ${\bf A}$ and ${\bf B}$ a two-layer thick film was first stabilized 
(corresponding approximately to points labeled {\bf 1} in Figs. 2 and 3,
we will refer to these conditions as ${\bf {A_1}}$ and ${\bf {B_1}}$ respectively). 
The friction force $F_f$ as a function of time was calculated for different 
shear velocities, and the results are shown in the main panel of Fig. 4. 
We show the time dependence of $F_f$ for the points of lowest velocities in the 
bottom part of Fig. 4.
At ${\bf {A_1}}$ the mean friction 
force goes to a constant when $v$ is reduced, as SSM sets 
in, around $v\sim 0.05~\sigma/\tau$, 
which reproduces previous results in related models\cite{20,21}. 
The limiting  value of the friction force corresponds in our case to a 
friction coefficient $\mu\simeq 0.2$. At 
${\bf {B_1}}$  the friction force decreases linearly as  
$v$ is reduced. No SSM is observed in this case\cite{nota}. 
This is consistent with the fluidity of the lubricant even in 
this highly confined geometry, a fact recently observed in water\cite{7}. 
The linear dependence the friction force $F_f(v)$ on $v$ 
can be used to estimate an effective viscosity  $\eta^{\rm eff}$ 
for the film as $\eta^{\rm eff}\equiv F_f(v)d/v \simeq 0.7~\varepsilon \tau/\sigma^3$, 
that is within three times the bulk value obtained by an independent 
measurement of the bulk single particle diffusivity ($D^{\rm bulk}_B\simeq 0.022~\sigma^2/\tau$)
and use of the Stokes-Einstein relation. This remarkable fact is to be compared 
with the finding that viscosity of water confined to molecular thicknesses 
has been shown\cite{7} to be no more than a few times the bulk value, and it 
can also be traced back to the high value of diffusivity obtained in 
the static calculation at point {\bf 1} in Fig. 3, which is close to the 
bulk value. 

Since temperature and pressure are different at points {\bf A} and {\bf B}, an 
additional check of the proposed scenario is desirable. Therefore, 
we did additional simulations in which the behavior is expected to 
be reversed: we performed fixed $d$ simulations at values $d=4.8~ \sigma$
at {\bf A}, and  $d=3.7~ \sigma$
at {\bf B} (labeled {\bf 2} in Figs 2 and 3, we will refer to these 
conditions as ${\bf {A_2}}$ and ${\bf {B_2}}$ respectively. Note that constant load simulations 
at these values of $d$ would be unstable). We have obtained in fact a 
big reduction  of the friction force at ${\bf {A_2}}$ 
with respect to ${\bf {A_1}}$, with 
no SSM detectable even at the lowest velocities (see bottom of Fig. 4), 
and an increase of the friction force at ${\bf {B_2}}$ 
with respect to ${\bf {B_1}}$, where 
now SSM becomes observable at the lowest velocities. 


Summarizing, we have shown that materials with expansive freezing are expected
to have good boundary lubrication capabilities, essentially due to the fact
that they do not show a strong tendency to solidify as they are squeezed.
This finding provides a novel strategy in the design of materials
for boundary lubrication.
At the same time our results give a qualitative 
explanation for the good lubrication
properties and low viscosity of very thin films of water. They also 
predict that experiments performed at constant $d$, at the squeeze-out 
points (such as ${\bf {B_2}}$) will produce larger friction forces than constant 
load experiments, contrary to what is expected in normal fluids. 
Similar results are likely obtained in other materials 
with expansive freezing.

I thank C. A. Balseiro and E. Tosatti for critical reading of the manuscript, and
A. Borsinger for technical assistance. 
This work was financially supported by CONICET, Argentina. Partial support 
by Fundaci\'on Antorchas (Argentina) and Italian contracts INFM/PRA NANORUB, 
and MIUR/COFIN is also acknowledged.

\begin{figure}
\narrowtext
\epsfxsize=3.3truein
\vbox{\hskip 0.05truein
\epsffile{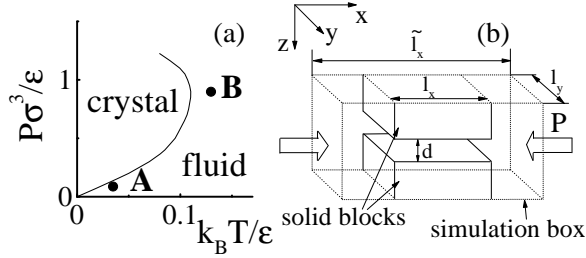}}
\medskip
\caption{(a) The melting line of the low pressure 
bulk crystalline structure of our model fluid. 
The points {\bf A} and {\bf B} are those simulated, with the results shown in 
Fig. 2 and 3 respectively. (b) A sketch of the geometrical configuration
used in the simulations. Fluid particles fill up the simulation box. Crystal
structure of solid blocks is not shown here. The external hydrostatic pressure
$P$ is indicated.}
\label{f0}
\end{figure}

\begin{figure}
\narrowtext
\epsfxsize=3.3truein
\vbox{\hskip 0.05truein
\epsffile{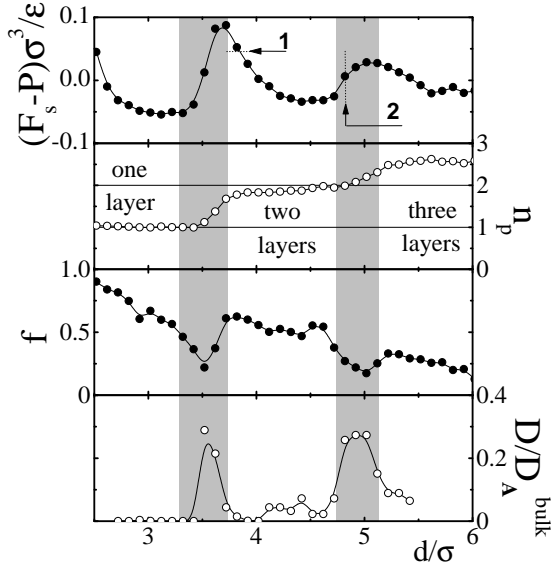}}
\medskip
\caption{Structure of the lubricant film at point {\bf A} of the phase diagram
as a function of the gap 
width $d$.
We show the equilibrium values of the solvation force $F_s$, 
the number of particles in the gap $n_p$
(normalized to a full layer of particles), 
the structure factor $f$ of the lubricant layers adjacent 
to the solids, and the horizontal diffusivity 
$D$ of particles in the gap ($D$ is measured relative to the bulk
three dimensional value at point {\bf A}, $D^{\rm bulk}_A\simeq 0.012~\sigma^2/\tau$).
In the gray regions there are drops 
of $F_s$, associated with the expulsion of a whole lubricant layer, as seen by
the evolution of $n_p$. This expulsion 
is correlated with a decrease of $f$ and an increase of $D$, which indicates 
that the film becomes more disordered, or fluid-like, in these regions.}
\label{f1}
\end{figure}

\begin{figure}
\narrowtext
\epsfxsize=3.3truein
\vbox{\hskip 0.05truein
\epsffile{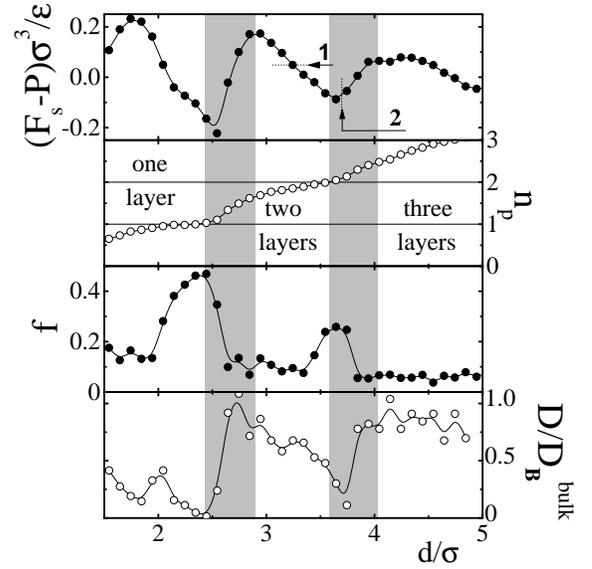}}
\medskip
\caption{Same as Fig.2 but at
point {\bf B} of the phase diagram. The 
trends for $F_s$ and $n_p$ are the same as those seen in Fig 2, but now the 
expulsion of a lubricant layer (within the gray regions) produces an 
ordering of the film, as seen from the increase of $f$ and the decrease of 
$D$ ($D^{\rm bulk}_B\simeq 0.022~\sigma^2/\tau$).}
\label{f2}
\end{figure}

\begin{figure}
\narrowtext
\epsfxsize=3.3truein
\vbox{\hskip 0.05truein
\epsffile{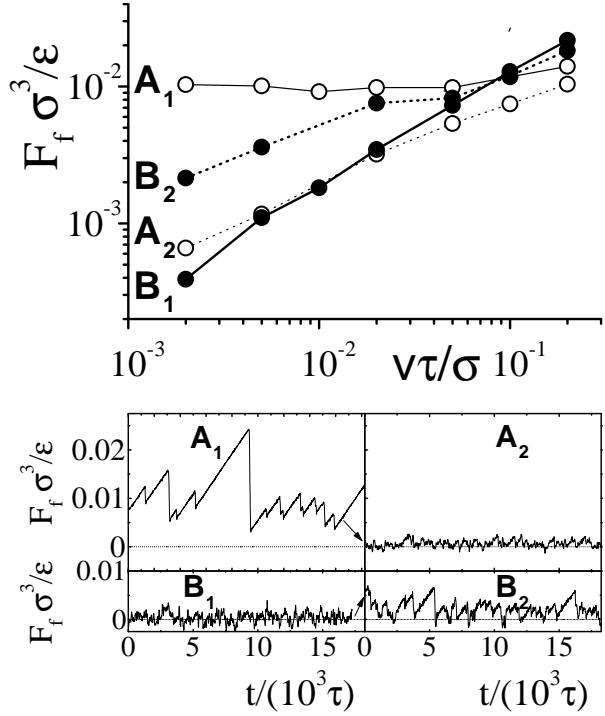}}
\medskip
\caption{Top: friction force per unit area $F_f$ as a function of shear velocity 
for points {\bf A} and {\bf B} of the phase diagram.
Continuous lines are results at a constant load $L$ of $0.05~\sigma^3/\varepsilon$ 
in excess of the hydrostatic pressure (points labeled {\bf 1} in Figs. 2 and 3).
Dotted lines are the results 
of simulations at constant 
$d$, chosen in such a way that now the film is quasi-liquid at ${\bf {A_2}}$, 
and quasi-solid at ${\bf {B_2}}$ (point labeled {\bf 2} in Figs. 2 and 3). 
Bottom: $F_f$  as a function of time
for the points of lowest velocities, displaying SSM at 
${\bf {A_1}}$ (for which $F_f$  goes 
to a finite value as $v$ is reduced) and uniform motion at ${\bf {B_1}}$ 
(for which $F_f$ decreases 
linearly as $v$ is reduced).  SSM
is strongly suppressed at ${\bf {A_2}}$ as compared to ${\bf {A_1}}$,  
and reappears in ${\bf {B_2}}$, as compared to ${\bf {B_1}}$.}
\label{f3}
\end{figure}

\end{document}